\documentstyle[twoside,fleqn,espcrc2]{article}

\newcommand{\AmS}{{\protect\the\textfont2
  A\kern-.1667em\lower.5ex\hbox{M}\kern-.125emS}}

\title{Disconnected Loop Noise Methods in Lattice QCD
\thanks{Presented at \lq\lq Lattice 97", Edinburgh, Scotland.}}

\author{Walter Wilcox and Bruce Lindsay\address{Department of Physics, Baylor 
								University, Waco, TX 76798-7316}}
       
\begin{document}

\begin{abstract}
A comparison of the noise variance between algorithms 
for calculating disconnected loop
signals in lattice QCD is carried out. 
The methods considered are the $Z(N)$ noise method and the
Volume method. We find that the noise variance is strongly influenced
by the Dirac structure of the operator.

\end{abstract}

\maketitle

\section{Introduction and Review}

The two most widely used numerical techniques for evaluating
disconnected amplitudes are the so-called $Z(N)$\cite{Thron,Liu}
 and Volume\cite{Japan} noise
methods. Disconnected diagrams are needed in calculations of, e.g.,
the pi-N sigma term (${\bar\psi}(x)\psi(x)$), 
nucleon quark spin content (${\bar\psi}(x)\gamma_{5}
\gamma_{\mu}\psi(x)$), pi-NN coupling (${\bar\psi}(x)\gamma_{5}\psi(x)$) and pion
polarizabilty. These two methods can be thought of as simply using
different noise vectors ($Z(N)$ or $SU(3)$) to evaluate summed, disconnected graph
amplitudes. Of course the Volume method is specific to gauge theories because
it utilizes Elitzur's theorem. Volume noises are simply equivalent to performing
random $SU(3)$ gauge transformations on the configuration. 

There are two sources of noise in the simulation. Given $N$ configurations
and $M$ random $Z(N)$ or Volume noises, the final error bar is given by
\begin{equation}
\sigma = \sqrt{\frac{S^{2}_{noise}}{NM}+\frac{S^{2}_{guage}}{N}}.
\label{error}
\end{equation}
The comparison here is between the $Z(N)$ and Volume values of $S^{2}_{noise}$
for various lattice operators. Such a comparison does not tell how many
configurations are necessary to aquire a signal for a given operator, but instead
which method will do the best job for a given amount of computer time.

The $Z(N)$ method is a general technique for inverting matrices, based upon
the solution of the system of equations,
\begin{equation}
Mx = b,
\label{Mx=b}
\end{equation}
where $M$ is the $N\times N$ quark matrix, $x$ is the solution vector and 
$b$ is the noise vector. It has the properties
\begin{equation}
<b_{i}>=0, <b_{i}b_{j}>=\delta_{ij},
\label{noise}
\end{equation}
where one is averaging over the noise vectors. Any inverse matrix element,
$S_{ij}=M^{-1}_{ij}$, can then be obtained from
\begin{equation}
<b_{j}x_{i}>= \sum_{k}S_{ik}<b_{j}b_{k}>=S_{ij}.
\label{Z(N)solution}
\end{equation}
We consider $Z(N)$ noise here, specifically the 
$Z(2)$ and $Z(4)$ noises. The reason for considering these two cases
is that it is known that the variance of any matrix element is the
same for $N\ge 3$, but in general the $N=2$ and $N\ge 3$ variances are
different\cite{Thron}. 

The difference in the variance of a given linear
combination
\begin{equation}
\sum_{ji}d_{ji}<b_{j}x_{i}>=\sum_{ij}d_{ij}S_{ij},
\label{linearn}
\end{equation}
between $Z(2)$ and $Z(N)$ ($N\ge 3$) is given by
\begin{eqnarray}
\lefteqn{Var[Z(2)]-Var[Z(N)]=} \label{complicated}\\ \nonumber
& & \frac{1}{L}\sum_{m,n,p,r;n\ne r}d^{T}_{nm}S_{mr}d^{\dagger}
_{rp}S^{*}_{pn}.
\end{eqnarray}
($L$ is the number of noise vectors.)
We have been able to find only a single local
operator ${\bar\psi}(x)C\psi(x)$ ($C$ is the charge conjugation
matrix; $C=\gamma_{2}$ here) for which the sign of the right-hand
side of Eq.(\ref{complicated}) can be predicted, imlpying 
that $Z(2)$ variance is smaller than $Z(N)$. However, 
we can not say how large this
difference is and a numerical simulation is needed.

An important issue in calculating the noise variance for the Volume method
is the generation of random SU(3) matrices. This can be done by calculating
the Haar measure for some given parametrization, but there is a simpler way.
Consider the mod of SU(3) by a copy of SU(2):
\begin{equation}
\frac{SU(3)}{SU(2)} \sim \frac{SO(6)}{SO(5)} \sim S^{5},
\label{coset}
\end{equation}
The coset space thus
consists of all points on a 5-sphere in 6 dimensions and the manifold of SU(3)
can be taken to be a 3-sphere (the SU(2) part) times a 5-sphere, giving 5+3=8 real
parameters. This gives a particularly easy way of generating random SU(3)
matrices. After establishing the relationship between the 5-sphere manifold and
the SU(3) parameter space, one need only choose random points on the spheres to
generate randomly distributed SU(3) matrices. We have found a parametrization that
holds everywhere on the 5-sphere except on a set of measure zero, namely a circle.

Another important issue is the fact that noise methods can be 
implemented by smearing over different subspaces. For example,
the Volume method is usually implemented by smearing over space-time
only, which means that $3\times 4=12$ inverses are necessary
to extract a complete column. (One could smear over
color indices as well using Elitzur's theorem, but we find no
computational advantage to doing so.) One can implement the $Z(N)$ noise
also with various smearings. The most common is to smear over all indices,
resulting in a complete column after a single inversion.
We will refer to a method which requires 12 input noise vectors to
give a single column as a \lq\lq 12 noise method", and a method which
needs only a single noise vector as a \lq\lq 1 noise method".

In carrying out our comparisons we consider all local operators,
${\bar\psi}(x)\Gamma\psi(x)$, as well as point-split versions of the
vector and axial vector operators. For each operator there are
both real and imaginary parts, but the quark propagator
identity
$
S = \gamma_{5}S^{\dagger}\gamma_{5},
\label{identity}
$
means that only the real {\it or} the imaginary part of each local or nonlocal
operator is nonzero on a given configuration. Our operators
are (averaged over all space-time points): \\ \\
\noindent
{\bf Scalar:}  $Re[{\bar\psi}(x)\psi(x)]$,\\
{\bf Local Vector:}   $Im[{\bar\psi}(x)\gamma_{\mu}\psi(x)]$,\\
{\bf P-S Vector:}  $\kappa Im[{\bar\psi}(x+a_{\mu})(1+\gamma_{\mu})
U^{\dagger}_{\mu}(x)\psi(x)-
{\bar\psi}(x)(1-\gamma_{\mu})U_{\mu}(x)\psi(x+a_{\mu})]$,\\
{\bf Pseudoscalar:}  $Re[{\bar\psi}(x)\gamma_{5}\psi(x)]$,\\ 
{\bf Local Axial:}  $Re[{\bar\psi}(x)\gamma_{5}\gamma_{\mu}\psi(x)]$,\\ 
{\bf P-S Axial:} $\kappa Re[{\bar\psi}(x+a_{\mu})\gamma_{5}\gamma_{\mu}
U^{\dagger}_{\mu}(x)\psi(x)+
{\bar\psi}(x)\gamma_{5}\gamma_{\mu}U_{\mu}(x)\psi(x+a_{\mu})]$,\\
{\bf Tensor:} $Im[{\bar\psi}(x)\sigma_{\mu\nu}\psi(x)]$.

\begin{table*}[hbt]
\setlength{\tabcolsep}{1.5pc}
\newlength{\digitwidth} \settowidth{\digitwidth}{\rm 0}
\catcode`?=\active \def?{\kern\digitwidth}
\caption{Pseudoefficiency (\lq\lq PE") of 12 noise methods vs.\ 1 noise $Z(2)$.}
\label{numbers}
\begin{tabular*}{\textwidth}{@{}l@{\extracolsep{\fill}}ll}
\hline
   &{PE($\frac{\rm{12\ noise\ Volume}}{\rm{1\ noise\ Z(2)}}$)} 
   &{PE($\frac{\rm{12\ noise\ Z(2)}}{\rm{1\ noise\ Z(2)}}$)} \\

\hline
{\bf Scalar:}    & $ 0.121E+02\, \pm\, 0.32E+01$ & $0.109E+02\, \pm\, 0.25E+01$ \\
{\bf Local Vector1:} & $0.953E+01\, \pm\, 0.21E+01$ & $0.871E+01\, \pm\, 0.18E+01$ \\
{\bf Local Vector2:} & $0.118E+02\, \pm\, 0.16E+01$ & $0.121E+02\, \pm\, 0.29E+01$ \\
{\bf Local Vector3:} & $0.958E+01\, \pm\, 0.25E+01$ & $0.114E+02\, \pm\, 0.24E+01$ \\
{\bf Local Vector4:} & $0.137E+02\, \pm\, 0.32E+01$ & $0.163E+02\, \pm\, 0.34E+01$ \\
{\bf P-S Vector1:}      & $0.121E+02\, \pm\, 0.30E+01$ & $0.994E+01\, \pm\, 0.22E+01$ \\
{\bf P-S Vector2:}      & $0.127E+02\, \pm\, 0.21E+01$ & $0.110E+02\, \pm\, 0.23E+01$ \\
{\bf P-S Vector3:}      & $0.968E+01\, \pm\, 0.16E+01$ & $0.114E+02\, \pm\, 0.15E+01$ \\
{\bf P-S Vector4:}      & $0.150E+02\, \pm\, 0.32E+01$ & $0.155E+02\, \pm\, 0.42E+01$ \\
{\bf Pseudoscalar:}     & $0.142E-01\, \pm\, 0.34E-02$ & $0.201E-01\, \pm\, 0.43E-02$ \\
{\bf Local Axial1:}     & $0.162E+00\, \pm\, 0.30E-01$ & $0.144E+00\, \pm\, 0.29E-01$ \\
{\bf Local Axial2:}     & $0.178E+00\, \pm\, 0.45E-01$ & $0.146E+00\, \pm\, 0.37E-01$ \\
{\bf Local Axial3:}     & $0.155E+00\, \pm\, 0.43E-01$ & $0.162E+00\, \pm\, 0.38E-01$ \\
{\bf Local Axial4:}     & $0.204E+00\, \pm\, 0.41E-01$ & $0.187E+00\, \pm\, 0.35E-01$ \\
{\bf P-S Axial1:}  & $0.183E+00\, \pm\, 0.31E-01$ & $0.151E+00\, \pm\, 0.18E-01$ \\
{\bf P-S Axial2:}  & $0.142E+00\, \pm\, 0.29E-01$ & $0.114E+00\, \pm\, 0.28E-01$ \\
{\bf P-S Axial3:}  & $0.197E+00\, \pm\, 0.60E-01$ & $0.186E+00\, \pm\, 0.49E-01$ \\
{\bf P-S Axial4:}  & $0.224E+00\, \pm\, 0.44E-01$ & $0.238E+00\, \pm\, 0.36E-01$ \\
{\bf Tensor41:} & $0.287E+00\, \pm\, 0.50E-01$  &   $0.295E+00\, \pm\, 0.49E-01$  \\
{\bf Tensor42:} & $0.128E+00\, \pm\, 0.26E-01$  &   $0.889E-01\, \pm\, 0.11E-01$  \\
{\bf Tensor43:} & $0.327E+00\, \pm\, 0.91E-01$  &   $0.398E+00\, \pm\, 0.13E+00$  \\
{\bf Tensor12:} & $0.471E+00\, \pm\, 0.64E-01$  &   $0.376E+00\, \pm\, 0.66E-01$  \\
{\bf Tensor13:} & $0.477E+00\, \pm\, 0.14E+00$  &   $0.363E+00\, \pm\, 0.53E-01$  \\
{\bf Tensor23:} & $0.562E-01\, \pm\, 0.91E-02$  &   $0.751E-01\, \pm\, 0.16 E-01$ \\
\hline

\end{tabular*}
\end{table*}

\section{Results and Conclusions}

The sample variance in $M$ quantities $x_{i}$ is given by
the standard expression:
\begin{equation}
S^{2}_{noise} = \frac{1}{M-1}\sum_{i=1}^{M}(x_{i}-{\bar x})^{2}.
\label{variance}
\end{equation}
What we concentrate on here are the {\it relative} variances between
the different methods. Since the square of the noise error is given by
$
\sigma_{noise}^{2}= S^{2}_{noise}/M,
\label{error}
$
a ratio of variances (weighted by the number of inverses or
noises required) gives a multiplicative measure 
of the relative efficiency. One important
caveat, however. We are doing a fixed number of iterations
for all of the operators; it could very well be that different
methods will require significantly different numbers of conjugate-gradient 
iterations to reach the same level of accuracy.
For these reasons, we prefer to
refer to our results as \lq\lq pseudo-efficiencies" (\lq\lq PE")
which are defined by
\begin{eqnarray}
{\rm PE(\frac{method1}{method2})}
\equiv\frac{N_{method1}(S^{2}_{noise1})^{method1}}
{N_{method2}(S^{2}_{noise2})^{method2}},
\label{Peff}
\end{eqnarray}
where $N_{method}$ are the number of noise
vectors required to achieve one column of the
inverse.

At present, we have results only for one rather small $\kappa$ value,
$\kappa=0.148$, using $10$ noises on $10$ configurations.
We look at 4 methods: $Z(2)$ (1 noise), $Z(4)$ (1 noise), $Z(2)$
(12 noise), and Volume (12 noise). Note that no gauge fixing
on the configurations was done in the $Z(N)$ noise cases.

A selection of our numerical results appears in Table 1
where only the pseudoefficiencies of the two
12 noise methods relative to 1 noise $Z(2)$ is presented. In this table
\lq\lq Local Vector1" means the local operator ${\bar\psi}(x)
\gamma_{1}\psi(x)$, \lq\lq P-S Vector1"
means the point-split version, and the local operator ${\bar\psi}(x)\gamma_{4}
\gamma_{1}\psi(x)$ is denoted as \lq\lq Tensor41", etc. A quick examination
of this table reveals that there are large and dramatic differences
in the pseudoefficiencies among the various operators.\ 1 noise methods 
are approximately 10 to 15 times more efficient than either 12 noise method
for scalar and vector operators; conversely, 12
noise methods are much more efficient for pseudoscalar, axial and tensor 
operators. The most extreme of the
entries is for the pseudoscalar, which is approximately
$60$ times more efficiently calculated with a 12 noise method
than with 1 column $Z(2)$. Although the results are not shown here, we have
also found pseudoefficiencies close to unity
for 12 noise Volume vs.\ 12 noise $Z(2)$ or of 1 noise $Z(2)$ vs.\ 1 noise $Z(4)$.

In conclusion, we have seen that the important distinction is
not between $Z(N)$ and Volume methods, but between 12 and 1 noise methods
and that the pseudoefficiencies are
strongly influenced by the Dirac structure of the operator.

This work is supported in part by NSF Grant No.\ 9722073 and the National Center
for Supercomputing Applications and utlized the SGI Power Challenge and
Origin 2000 systems. We gratefully acknowledge conversations
regarding the coset space of SU(3) with Ronald Stanke.

\end{document}